**FULL TITLE:**

# Non-invasive Reversible Software-based Electron FLASH Irradiation Configuration of a Linear Accelerator in Clinical Use

**RUNNING TITLE:**

Non-invasive FLASH Configuration of Clinical TrueBeam

**AUTHOR NAMES:**

Stavros Melemenidis, PhD[1*,†], Dixin Chen, MS[1*], Cody Jensen[1,††], Joseph B. Schulz, BS[1,†††], Murat Surucu, PhD[1], Amy S. Yu, PhD[1], Edward E. Graves, PhD[1], Mengying Shi, PhD[2], Michael Reilly, PhD[2], Peter G. Maxim, PhD[2], Andrew Currell[3], Billy W. Loo Jr, MD, PhD[1], Lawrie Skinner, PhD [1], and M. Ramish Ashraf, PhD[1].

**AFFILIATIONS:**

[1] Department of Radiation Oncology, Stanford University School of Medicine, Stanford, CA 94305, USA

[2] Chao Family Comprehensive Cancer Center, University of California Irvine, Irvine, CA 92697, USA

[3] Accelerator Service, Department of Medical Physics, Princess Margaret Cancer Centre, University Health Network, Toronto, ON, Canada

**CURRENT ADDRESS:**

[†] Department of Radiation Oncology, Anschutz Medical Campus, University of Colorado, Aurora, CO 80045, USA

[††] Department of Civil and Environmental Engineering, Samueli School of Engineering, University of California, Irvine, CA 92697, USA

[†††] Department of Medical Physics, University of Wisconsin–Madison, Madison, WI 53705, USA

**CO-FIRST AUTHORS:**

Stavros Melemenidis and Dixin Chen

**CO-SENIOR AUTHORS:**

Billy W. Loo Jr, Lawrie Skinner, and M. Ramish Ashraf

**CORRESPONDING SENIOR AUTHORS:**

M. Ramish Ashraf
Department of Radiation Oncology
Stanford University School of Medicine
Telephone: +1 (650) 723-5591
Email: ramish.ashraf@stanford.edu

Lawrie Skinner
Department of Radiation Oncology
Stanford University School of Medicine
Telephone: +1 (650) 723-1412
Email: lawries@stanford.edu

Billy W. Loo, Jr.
Department of Radiation Oncology
Stanford University School of Medicine
Telephone: +1 (650) 736-7143
Email: bwloo@stanford.edu

**AUTHOR FOR EDITORIAL CORRESPONDENCE:**
Dixin Chen
Department of Radiation Oncology,
Stanford University School of Medicine
Stanford, CA 94305
Email: dixin@stanford.edu

**AUTHOR CONTRIBUTIONS:** All authors met the *International Committee of Medical Journal Editors (ICMJE) criteria for authorship*. Stavros Melemenidis and Dixin Chen are the co-first authors, each of whom is appropriately listed first when referencing this publication on their respective CVs. M. Ramish Ashraf, Lawrie Skinner, and Billy W. Loo Jr are the co-senior/corresponding authors.

**FUNDING:** This study was supported by funding from NIH grants P01CA244091 (EEG, PGM, BWL), R01CA26667 (PGM, EEG, BWL), AAPM-ASTRO Trainee Seed Grant (MRA). We also gratefully acknowledge philanthropic donors to the Department of Radiation Oncology at Stanford University School of Medicine.

**ACKNOWLEDGEMENTS:** We would like to thank Riccardo Dal Bello, Stephanie Tanadini-Lang, Matthias Guckenberger, and Daniel Letourneau for helpful discussions.

**DISCLOSURES:** B.W.L. is a co-founder and board member of TibaRay. B.W.L. has received lecture honoraria from Mevion. P.G.M. is a co-founder and board member of TibaRay. All other authors declare no conflicts of interest.

**Data Availability**: Research data are stored in an institutional repository and will be shared upon request to the corresponding author.

## Abstract

**Background:** Configuring clinical linear accelerators (linacs) for ultra-high dose rate (UHDR) electron experiments typically requires invasive hardware manipulation and/or irreversible modifications, limiting broader implementation. This work reports a reversible and non-invasive UHDR electron configuration of a clinical TrueBeam linac that enables switching between preclinical UHDR and conventional clinical treatment (CONV) modes through software settings, without accessing the linac interior.

**Methods:** Built-in service mode software was used to configure the UHDR mode with settings typical of a standard megavoltage photon beam. Using service mode, the photon target and monitor chambers were retracted. An unused low-energy electron scattering foil was loaded. An external AC current transformer for beam control and monitoring was mounted on the accessory tray, with an ionization chamber placed downstream in solid water to monitor exit dose. Dose profiles were measured for UHDR and CONV beams with radiochromic films for open field, *in vivo*, and *in vitro* setups. Dose-per-pulse (DPP) was varied by adjusting the gun voltage and quantified. Day-to-day output variation was assessed to evaluate dose reproducibility.

**Results:** Percent Depth Dose measurements confirmed similar energy between UHDR (9.2 & 12.6 MeV) and CONV electron beams (8.4 & 11.7 MeV), with matching profiles throughout the typical thickness of a mouse or cell culture media. Maximum DPP reached 1.5 Gy/pulse and 0.7 Gy/pulse for *in vivo* and *in vitro* setups at 64 cm and 82 cm source-to-surface distances respectively. Field flatness and symmetry were maintained between UHDR and CONV, supporting organ-specific *in vivo* irradiation and a maximum of 5 × 5 $cm^2$ field for *in vitro* irradiation. Day-to-day output variation remained small, with both inter- and intra-animal coefficients of variation averaging <3% for FLASH and <1% for CONV.

**Conclusion:** Accurate and reproducible UHDR electron delivery was demonstrated without invasive hardware manipulation, enabling preclinical FLASH research on a clinical linac.

## Introduction

Ultra-high-dose-rate (UHDR) radiotherapy, also known as FLASH, delivers ultra-rapid irradiation (e.g., dose rates exceeding 40–100 Gy/s), considerably faster than conventional radiotherapy (~0.1 Gy/s). A growing amount of preclinical evidence suggests that FLASH irradiation can substantially reduce normal tissue toxicity while maintaining tumor control, motivating intense interest in its biological mechanisms and clinical translation [1-4]. Despite increasing interest, access to UHDR irradiation platforms remains limited, as most FLASH studies rely on prototype accelerators, research-only linacs, or decommissioned clinical machines requiring invasive or nongeneralizable modifications [2, 5-7].

Electron beams remain the most accessible modality for FLASH research due to their relative ease of adaptation to UHDR delivery [8-17]. Previous configurations of clinical Varian linacs capable of UHDR delivery have required physical modification of the treatment head, most commonly by operating under photon mode with the X-ray target and flattening filter removed to achieve sufficiently high electron dose rates [8, 9, 13-15]. As a result, these approaches have typically been implemented on decommissioned systems, limiting their integration into routine clinical environments. Although one study achieved software-only UHDR delivery on a Varian TrueBeam system, it relied on proprietary software or non-public modifications, limiting its accessibility [12].

This brief report demonstrates that a TrueBeam linac in routine clinical use can be configured non-invasively and reversibly for UHDR electron irradiation using user-accessible service-mode settings and external beam monitoring. This study then characterizes beam energy, spatial uniformity, dose-per-pulse (DPP), pulse-to-pulse stability, and dosimetric equivalence with conventional electron beams across open field, representative *in vivo*, and *in vitro* preclinical setups. Lastly, this study further assesses day-to-day dose reproducibility during multi-fraction whole-brain irradiation. These results establish a practical and reproducible UHDR electron platform on a widely available clinical linac, enabling rigorous preclinical FLASH radiobiology studies without disruption to clinical operation. Notably, while prior implementations on Elekta systems have similarly required hardware modification [9, 18], our approach suggests a pathway by which such platforms could also be adapted for UHDR delivery and beam control optimization using software-only configurations.

## Materials and Methods

### UHDR Configuration

A clinically commissioned Varian TrueBeam linac was configured non-invasively and reversibly for UHDR electron irradiation using service-mode software. Figure **1** showed experimental setup and beam monitoring configuration for UHDR on the linac. The configuration employed gun and radiofrequency (RF) settings corresponding to MV photon operation, with software-controlled retraction of the photon target and internal monitor chambers, and insertion of a clinically unused 4 MeV electron scattering foil (Figure **1**a). UHDR beam monitoring and control were achieved using an external gating and monitoring system interfaced with the linac (Figure **1**b, c). A high-speed relay, controlled via a LabVIEW program, enabled or inhibited beam delivery following RF stabilization. An external AC current transformer (ACCT) provided pulse-resolved measurements, which were used for beam monitoring and to terminate delivery after a predefined number of pulses. The total dose was set by the number of pulses, while the dose per pulse was adjusted via the gun grid voltage, with calibration factors used for dose estimation. No physical access to the linac interior or permanent hardware modification was required. Full configuration details are provided in Supplementary Sections S**1**-**4**.

Representative preclinical setups included an *in vivo* whole-brain irradiation setup using a 1.7 cm diameter circular Cerrobend collimator and a stereotactic mouse holder at 64 cm source-to-surface distance (SSD), and an *in vitro* cell-culture setup with no collimation at 82 cm SSD (Figures **1**d, e). Percent depth dose (PDD), lateral profiles, and DPP were measured using EBT-XD radiochromic film. An Advanced Markus chamber positioned downstream in solid water was used to provide additional dosimetric reference for both UHDR and CONV beams. Day-to-day dose reproducibility was assessed during multi-fraction *in vivo* irradiation. Detailed dosimetric methods and analysis procedures are described in Supplementary Sections S**5**-**7**.

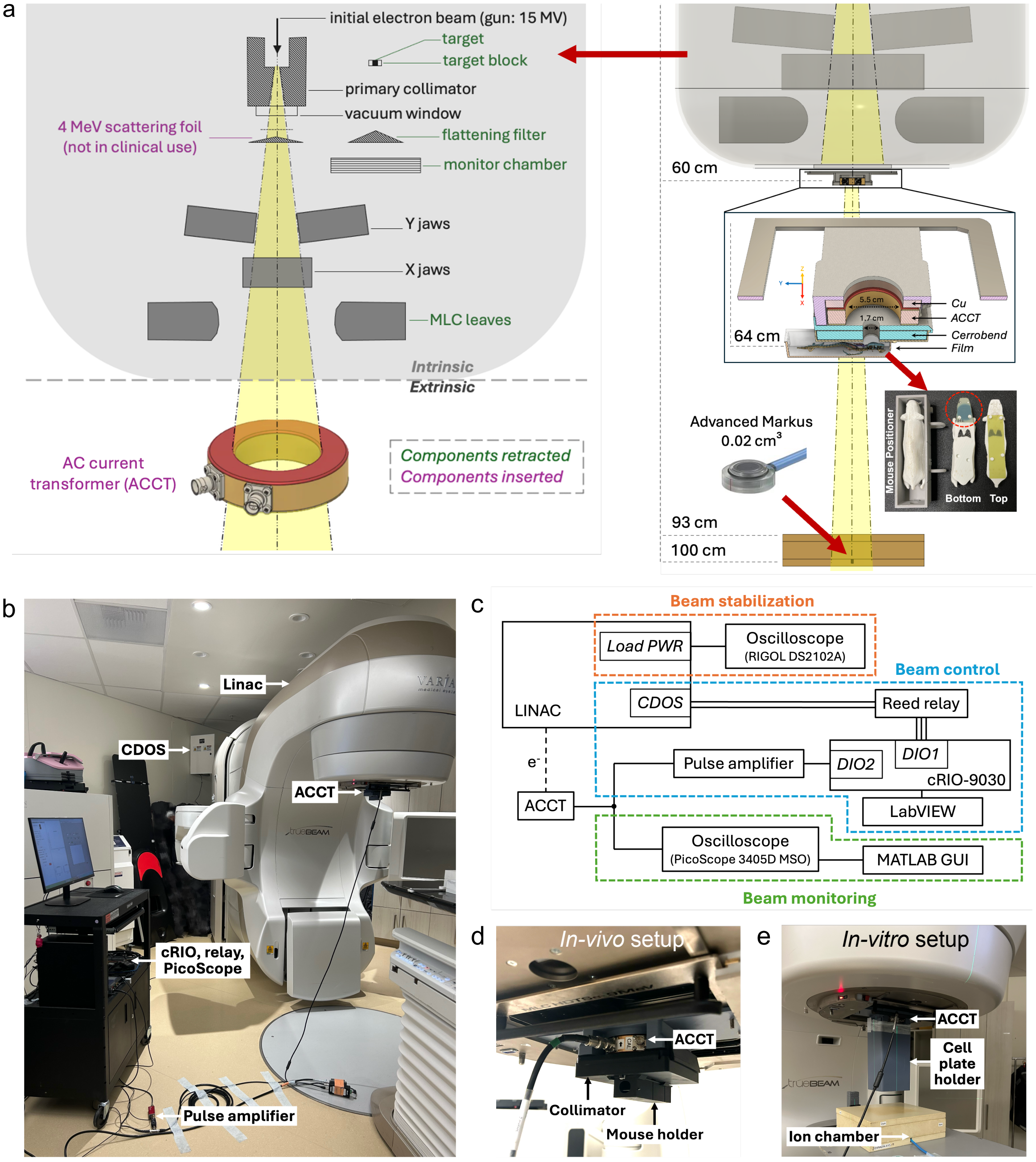

***Figure 1 Non-invasive FLASH setup on a linac. a.*** *Schematic (not to scale) of the FLASH configuration at the linac head. On the left, it shows components non-invasively retracted from the beam path (green text) and components introduced into the beam path (pink text). On the right, it shows the external FLASH unit mounted onto the linac head, including an ACCT with copper shield, a Cerrobend collimator, a jig with a mouse phantom and films placed inside, and an optional Advanced Markus chamber downstream in solid water slabs.* ***b.*** *Experimental setup in the treatment vault, showing the linac, ACCT, CDOS, cRIO controller, relay, PicoScope oscillator, and pulse amplifier used for beam control and monitoring.* ***c.*** *Schematic of the beam stabilization, control, and monitoring methods for FLASH delivery. Load PWR: reflected radiofrequency power in load cells; DIO1/2: digital input/output channels; CDOS: customer-defined dosimetry; GUI: graphic user interface. See Section S4 for more details.* ***d.*** *In-vivo experiment setup mounted to the linac head with an organ-specific collimator and a mouse holder.* ***e.*** *In-vitro setup mounted to the linac head with a cell plate holder and an optional ionization chamber downstream.*

## Beam Characterization

UHDR beams at two nominal energies, 9 MeV and 12 MeV, were delivered by operating the linac under 12 MV and 15 MV photon energy settings, respectively, with the target retracted remotely in the service mode. Corresponding CONV beams were delivered under clinical 9 MeV and 12 MeV electron settings. PDD and lateral profile measurements were performed at each nominal energy. The depth of the 50% isodose level ($R_{50}$) was determined from the PDDs and used to estimate the mean surface energy as $0.656 + 2.059 \times R_{50} + 0.022 \times R_{50}^2$ [19]. For the 12 MeV UHDR configuration, PDDs and dose profiles were measured under open field (100 cm SSD, no collimation), and the aforementioned *in vivo* (Figure **1**d) and *in vitro* (Figures **1**e and **S8**) setups. For the latter two, DPP was measured across a range of gun grid voltage settings. All dosimetry was performed using EBT-XD films, utilizing a custom water tank for open-field setup and a stack of solid water slabs for the *in vivo* and *in vitro* setups.

Pulse-to-pulse stability was characterized by acquiring the voltage waveform of each pulse using the ACCT under the *in vitro* setup (no collimation) with an 8 × 8 $cm^2$ jaw size. Two sets of five consecutive deliveries of 100 pulses at a gun voltage of 163 V were performed before and after machine warm-up, which consists of five 100-pulse deliveries at each of six gun grid voltages spanning 140-165 V. For each pulse, area under the waveform and pulse width calculated as full width at half-maximum were extracted. The area under the waveform was then converted to dose per pulse based on film measurements. Overall mean dose per pulse and pulse width were calculated, and the time required for the beam to reach to the mean was defined as the ramp-up time.

## Calibration and Day-to-day reproducibility

To evaluate the day-to-day reproducibility of dose delivery under both UHDR and CONV irradiations, data from a five-day fractionated whole-brain irradiation experiment were analyzed. Mice were assigned to UHDR or CONV irradiation groups and treated using identical 1.7 cm circular collimation and positioning at 64 cm SSD. Mice were irradiated with target doses of 5 or 6 Gy: 28 in the UHDR group ($n$ = 14 per dose, ~1 Gy/pulse) and 32 in the CONV group ($n$ = 16 per group). Each mouse received one fraction per day over five days.

Dose calibration was performed using a 3D-printed mouse phantom at 0.9 cm depth [20]. Prescribed doses were mapped to target ACCT charges or monitor units (MU) by pre-established calibration curves at a range of outputs. Quality assurance (QA) film measurements were taken on each experiment day. Delivered doses were calculated using *a posteriori* calibration curve derived from the daily QA films correlated with ACCT charges (UHDR) or ionization chamber readings (CONV). More calibration details are described in Supplementary Section S**8**.

For day-to-day reproducibility evaluation, delivered doses were calculated as a percentage of the target dose per fraction for both UHDR and CONV groups. Inter-animal coefficients of variation (CV, the ratio of the standard deviation to the sample mean) were calculated for each day to quantify group-level variability, and intra-animal CVs were computed across all five days to evaluate within-subject consistency.

# Results

## Beam Characterization

Figure **2** shows depth-dose measurements to assess beam penetration and dose distribution under the open field, *in vivo*, and *in vitro* setups. Beam characterization with open field at the isocenter (100 cm SSD) demonstrated a consistent shift in the PDD profiles between UHDR and CONV (Figures **2**a, b). For 9 MeV, the UHDR beam exhibited an $R_{50}$ of 4.0 cm, compared to 3.6 cm for the CONV beam, corresponding to mean surface energies of 9.2 MeV and 8.4 MeV respectively. A similar trend was observed for 12 MeV, where the $R_{50}$ was 5.5 cm for UHDR and 5.1 cm for CONV, corresponding to mean surface energies of 12.6 MeV and 11.7 MeV respectively.

Surface lateral profiles demonstrated high dose uniformity within the irradiation field for both UHDR and CONV beams (Figure **2**c). Across the full 20 × 20 $cm^2$ area, profiles along the $x$ and $y$ axes exhibited minimal deviation from the beam center. Within ±2 cm of the beam center, flatness for the UHDR beam was 5.8% along the *x*-axis and 5.3% along the *y*-axis, while the CONV beam showed flatness values of 4.2% (*x*) and 5.7% (*y*). Symmetry measurements within ±2 cm of the beam center were 4.6% (*x*) and 5.2% (*y*) for UHDR, and 2.2% (*x*) and 2.9% (*y*) for CONV.

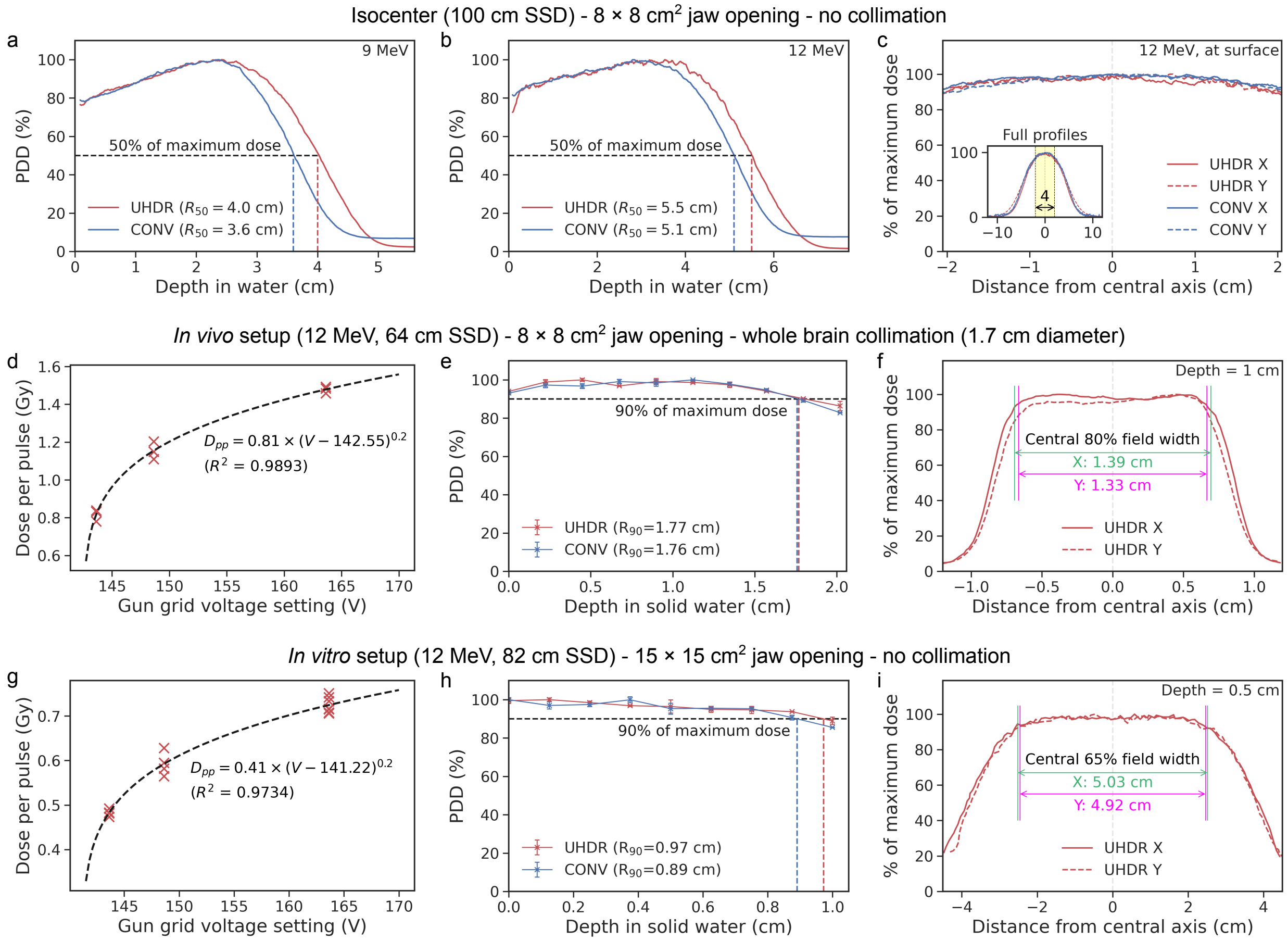

***Figure 2: Dosimetric characterization of UHDR and CONV beams across open field, in vivo, and in vitro setups. a, b.*** *Percentage depth dose (PDD) curves acquired using a custom water tank at 100 cm SSD with open field for the nominal 9 MeV and 12 MeV UHDR and CONV electron beams with dashed lines indicating the depth of the 50% isodose level (R50).* ***c.*** *Central 4 × 4 cm2 lateral x and y profiles for 12 MeV UHDR and CONV beams at 100 cm SSD with open field, with full 20 × 20 cm2 field profiles shown in the bottom-left corner.* ***d, g.*** *Dose per pulse as a function of gun grid voltage setting for 12 MeV UHDR beams with the in vivo and in vitro setups fitted to power law equations (dashed lines with fitted equations and the coefficients of determination R2 indicating the goodness of fit).* ***e, h.*** *PDD curves of the 12 MeV UHDR and CONV beams using in vivo and in vitro setups respectively. Dashed lines indicate the 90% isodose level.* ***f, i.*** *Lateral x and y profiles of the 12 MeV UHDR beam for the in vivo (measured at 1 cm depth) and in vitro (0.5 cm depth) setups. Double sided arrows denote the central 80% and 65% of field widths for in vivo and in vitro respectively, which are the regions used for flatness and symmetry calculations.*

The optimal gun grid voltages for UHDR delivery under the *in vivo* and *in vitro* setups were identified by evaluating the output in terms of dose per pulse across a range of gun grid voltage settings under fixed RF settings (Figures 2d, g). The output response curves were fitted to power-law equations, exhibiting a slower increase in output beyond 150 V. Although the power-law fits predicted increasing output at higher voltages, the measured dose per pulse saturated at 163.62 V. At this voltage, the *in vivo* setup (whole-brain collimator) and *in vitro* setup (uncollimated) yielded maximum doses per pulse of 1.48 Gy/pulse and 0.72 Gy/pulse respectively. The output response curve provides the necessary dynamic range to achieve the various instantaneous dose rates for FLASH experiments while maintaining the ability to deliver matched total CONV doses.

For the *in vivo* setup with brain collimator at 64 cm SSD, both 12 MeV UHDR and CONV beams demonstrated comparable PDD profiles within 2 cm of solid water, simulating the typical thickness of a mouse (Figure 2e). The dose plateaued within the first 1 cm and gradually declined, reaching approximately 90% of the maximum dose at a depth of 1.8 cm for both UHDR and CONV. At 1 cm depth, lateral profiles within the central 80% field width ($x$ = 1.4 cm, $y$ = 1.3 cm) exhibited flatness values of 4.8% ($x$) and 5.5% ($y$), and symmetry values of 2.3% ($x$) and 4.8% ($y$) (Figure 2f).

For the *in vitro* setup with no collimation at 82 cm SSD, the PDD measurements demonstrated comparable dose fall-off profiles for UHDR and CONV beams (Figure 2h). Within the 1 cm depth range analyzed, simulating typical culture media depths, dose deposition exceeded 90% of the maximum ($R_{90}$ = 1.0 cm for UHDR, 0.9 cm for CONV). Lateral dose profiles acquired confirmed sufficient spatial uniformity for UHDR delivery (Figure 2i). At 0.5 cm depth, lateral profiles within the central 65% field width ($x$ = 5.0 cm, $y$ = 4.9 cm) exhibited flatness values of 4.2% ($x$) and 4.3% ($y$), and symmetry values of 2.3% ($x$) and 4.2% ($y$).

Pulse-to-pulse stability was assessed across two sets of five 100-pulse deliveries before (Figures 3a, b) and after (Figures 3c, d) machine warm-up at a gun grid voltage of 163 V. Prior to warm up, the dose per pulse exhibited a transient ramp-up increasing 2-3% over the first 8 pulses (<44ms) and plateaued at an overall mean dose per pulse of 0.519 Gy/pulse (CV = 0.9%). In contrast, no ramp-up was observed in the pulse width (overall mean 3.78 µs, CV = 0.3%), indicating that the transient affects pulse amplitude but not temporal pulse structure. After warm-up, no ramp-up behavior was observed in either area under the waveform or pulse width for all deliveries, with overall mean dose per pulse of 0.518 Gy/pulse (CV = 0.8%) and overall mean pulse width of 3.79 μs (CV = 0.3%). The variation for each delivery was under 3% and 1% from the overall mean for the dose per pulse and pulse width respectively both before and after warm-up.

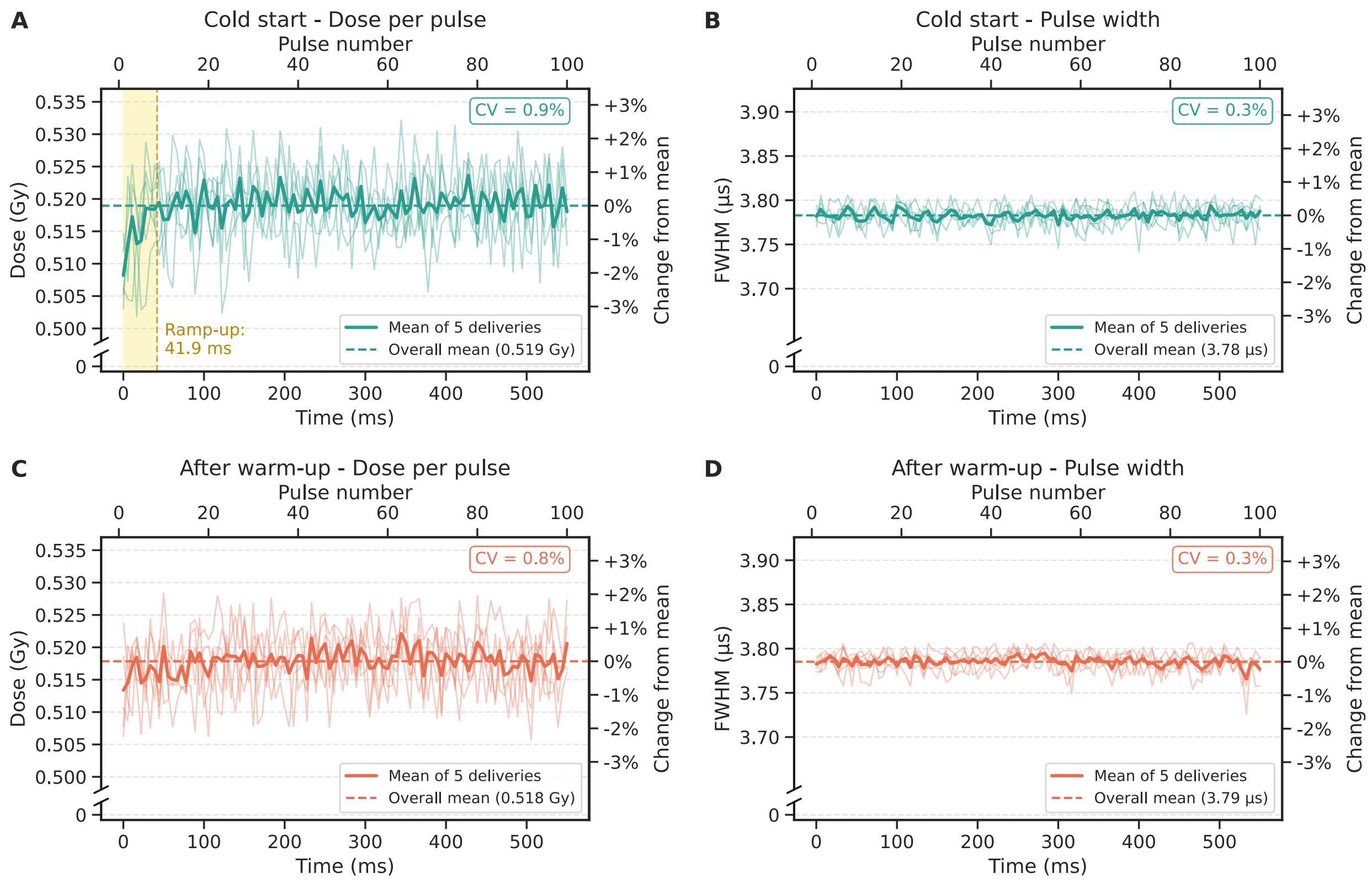

***Figure 3: Pulse-to-pulse stability evaluated using dose per pulse and pulse width (full width at half maximum, FWHM) over 100 pulses for five consecutive deliveries, before and after machine warm-up. a, b.*** *Cold-start deliveries show a ramp-up of 41.9 ms in the dose per pulse, while FWHM remains stable across 100 pulses.* ***c, d.*** *Post-warm-up deliveries show no observable transient. The variation from the overall mean is within 3% and 1% for the dose per pulse and pulse width respectively both before and after warm-up. Bold traces represent the mean of five deliveries and light traces show individual deliveries.*

## Day-to-day reproducibility

Day-to-day dose reproducibility was evaluated during five consecutive days of the fractionated whole-brain irradiation for both UHDR and CONV groups. Delivered dose normalized to the target dose per fraction remained tightly clustered around the intended prescription for both modalities (Figure 4). Detailed statistics for each day and averaged across each mouse are shown in Table 1. For UHDR, inter-animal CVs ranged from 2.2% to 3.3% across five days with an average value of 2.7%, and the average intra-animal CV was 2.7%. For CONV delivery, both average inter- and intra-animal CVs remained below 1%. Overall, dose delivery remained stable across all five treatment days and within acceptable experimental tolerances.

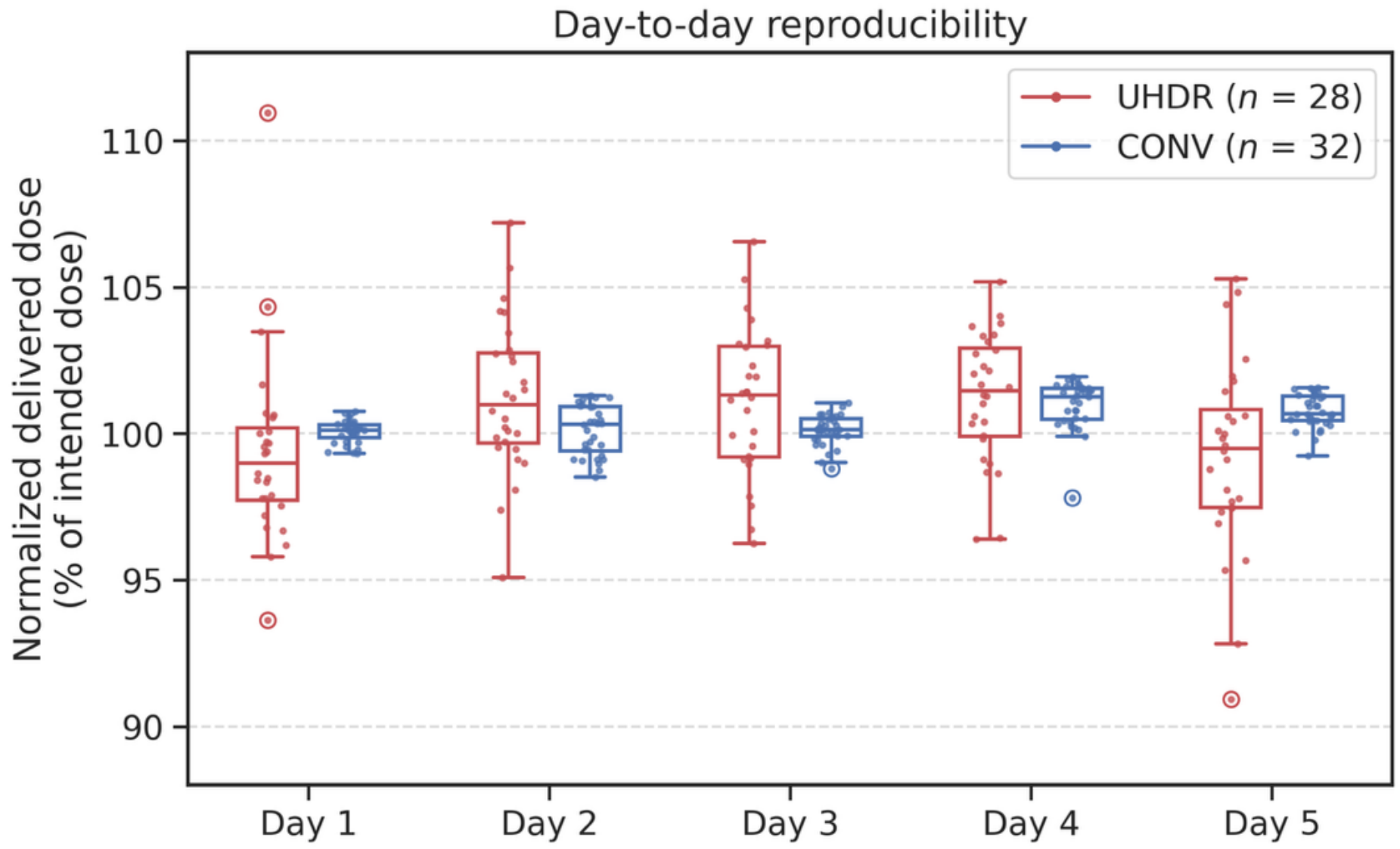

***Figure 4: Day-to-day reproducibility of dose delivery for UHDR and CONV beams.*** *Delivered dose as a percentage of the target dose over a five-day fractionated irradiation for UHDR (n = 28, red) and CONV (n = 32, blue). Boxes indicate the median and interquartile range, with whiskers indicating the daily variance and all data points overlaid for each daily measurement (dots inside an open circle represent the outliers).*

**Table 1**: Day-to-day reproducibility statistics for UHDR and CONV irradiations.

| | **UHDR** | | | **CONV** | | |
|---|---|---|---|---|---|---|
| **Statistics** | **Mean (%)** | **SD (%)** | **CV (%)** | **Mean (%)** | **SD (%)** | **CV (%)** |
| *Inter-animal* | | | | | | |
| Day 1 | 99.7 | 2.9 | 2.9 | 100.0 | 0.8 | 0.8 |
| Day 2 | 101.2 | 2.6 | 2.6 | 100.1 | 0.7 | 0.7 |
| Day 3 | 101.1 | 2.5 | 2.5 | 100.2 | 0.8 | 0.8 |
| Day 4 | 101.2 | 2.2 | 2.2 | 100.3 | 0.6 | 0.6 |
| Day 5 | 99.2 | 3.3 | 3.3 | 100.1 | 0.7 | 0.7 |
| **Mean** | – | – | **2.7** | – | – | **0.7** |
| *Intra-animal* | | | | | | |
| Minimum | – | – | 1.1 | – | – | 0.3 |
| Maximum | – | – | 6.2 | – | – | 1.2 |
| **Mean** | – | – | **2.7** | – | – | **0.8** |

SD: standard deviation; CV: coefficient of variation (100 × SD / mean).

### UHDR operational history

Over approximately 15 months of operation on this platform, the total number of pulses delivered in the UHDR mode is conservatively estimated to be 33,000 pulses, corresponding to a cumulative beam-on time of about 3 minutes assuming continuous operation at a repetition rate of 180 Hz. Throughout this period, no deviations in clinical QA metrics have been observed beyond the expected temporal drift.

## Discussion and Conclusions

This work demonstrates that a TrueBeam linac in routine clinical use can be configured non-invasively and reversibly for UHDR electron irradiation suitable for preclinical *in vivo* and *in vitro* UHDR radiobiological research. By controlling the retraction of the photon target and monitor chambers in the software, this platform achieves UHDR delivery without physical access to the linac interior or exposure of clinically used components to non-standard dose rates. This distinguishes the TrueBeam implementation from prior clinical linac configurations that relied primarily on electron beam modes or invasive modifications to achieve UHDR operation.

The system provides energy-matched UHDR and CONV electron beams with comparable depth dose characteristics and spatial uniformity, supporting radiobiological comparisons for FLASH studies. Achievable doses per pulse reached approximately 1.5 Gy/pulse for the collimated *in vivo* setup and approximately 0.7 Gy/pulse for the *in vitro* setup. Importantly, dose delivery was reproducible across multi-day fractionated experiments, with <3% inter- and intra-animal variability, meeting output stability requirement for robust preclinical FLASH studies.

While the present work focuses on preclinical applications, the widespread clinical availability of the TrueBeam platform and the fully reversible nature of this configuration lower barriers to broader adoption and multi-institutional investigation. With additional engineering refinements and appropriate clinical safeguards, this approach may also support future translation of UHDR electron delivery in early-phase clinical trials.

# Non-invasive FLASH Configuration of Clinical TrueBeam: Supplementary Information

## S1 Gating Configuration (one-time initial setup)

The following methods describe the configuration necessary for controlling the Customer-Defined Dosimetry (CDOS) interlock on a TrueBeam linear accelerator (linac).

### S1.1 CDOS Interlock

The CDOS interlock is a flexible framework provided by the manufacturer that allows users to define custom interlock conditions by tapping into the beam-on line. This capability enables customer-driven gating logic, where external signals can initiate or terminate beam delivery. In our implementation, the CDOS interlock was configured to interrupt the beam after a predefined number of pulses, allowing precise control of the total dose delivered.

### S1.2 Enabling CDOS Interlock in Advanced Service Mode

From Advanced Service Mode, from the External Interface settings, we accessed the dual in-line package (DIP) switches configuration for the beam enable loop (BEL) and set the DIP switch 4 (in our system) to the "ON" position (Figure S1). This action activated the CDOS interlock and automatically enabled a custom spare switch (switch 2 in our setup).

### S1.3 Hardware Configuration for CDOS Interlock

The hardware configuration required to enable the CDOS interlock on TrueBeam linacs depends on the specific model. After safely powering down the linac, we accessed the gantry stand on the appropriate side for our model. The BEL printed circuit board is shown in Figure S2, and DIP switch 4 (in our setup) was set to the "ON" position in alignment with the software settings described above.

### S1.4 Relay Junction Box (RJB) Configuration

To configure the CDOS interlock, we established a connection using a normally closed switch between the 24V line (terminal TB3-2) and both the CDOS (terminal TB3-1) and custom spare switch 2 (terminal TB3-7) inputs. Disconnecting the 24V line from these inputs activates the CDOS interlock. The exact location of the CDOS interlock varies by TrueBeam model. Older models typically have the CDOS interlock located on the right-hand side of the gantry, whereas newer models feature a Relay Junction Box (RJB) usually placed on a room wall or behind the gantry (Figure S3).

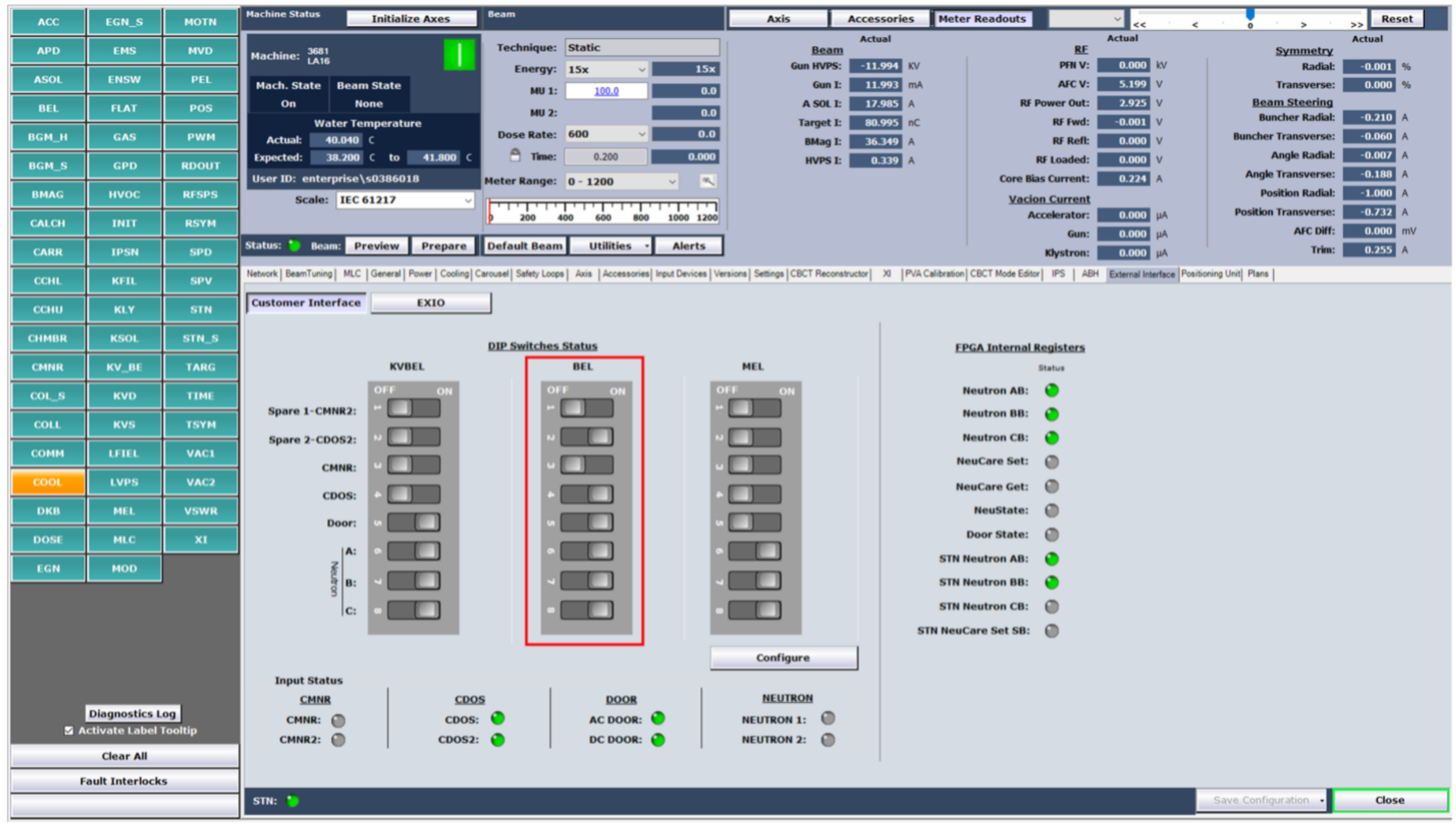

Figure S1: Beam enable loop (BEL) dual in-line package (DIP) switches configuration highlighted in a red box under External Interface tab in Advanced Service Mode for enabling the customer-defined dosimetry (CDOS) interlock.

## S2 Carousel Positioning and Component Retraction

The carousel coordinates can be configured from the Carousel tab while logged into Advanced Service Mode (Figure S4). The adjustable parameters include radial position, transverse position, ion chamber alignment, target alignment, and energy switch position.

### S2.1 Energy Switch

As a first step in configuring the system for FLASH, it is essential to ensure that the Energy Switch position in the carousel remains aligned with the selected photon energy used for the configuration. To avoid potential mismatches, it is recommended to first load the desired energy from the Beam box on the main screen of the Advanced Service Mode before accessing the Carousel tab. This action automatically sets the Energy Switch position on the carousel, which should remain unchanged as long as the same energy is maintained. This precaution is particularly important when switching between different energies (e.g., 15X mode for 12 MeV FLASH and 10X mode for 9 MeV FLASH), which have different Energy Switch positions and it can easily result in an Energy Switch position mismatch.

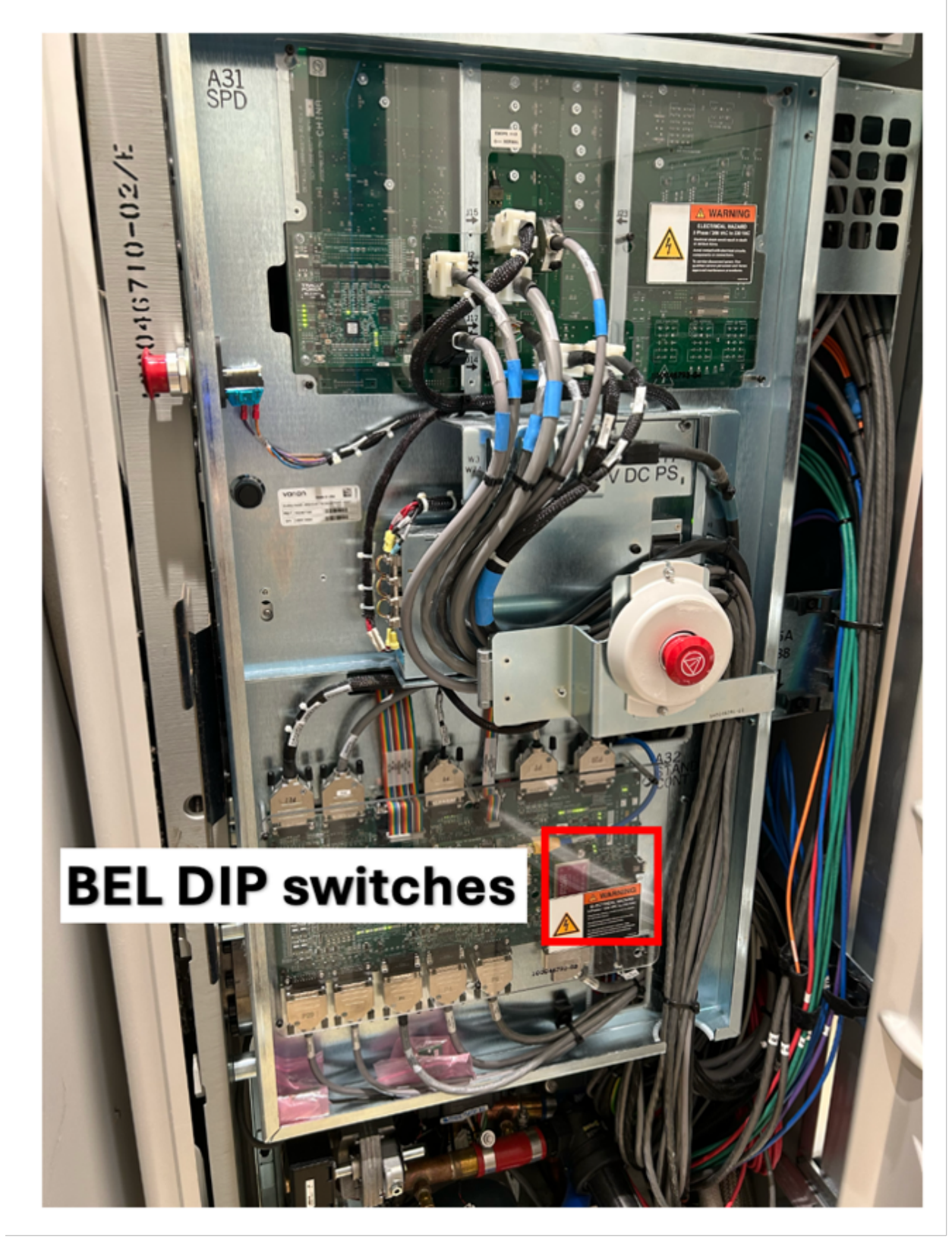

Figure S2: Location of the beam enable loop (BEL) dual in-line package (DIP) switches inside the TrueBeam gantry stand.

## S2.2 Determination of Coordinates for Spare Scattering Foil

To establish the coordinates for the spare scattering foil, we selected the 4 MeV port. This is not used in clinically in our center because we do not treat with 4 MeV electrons on this machine, and we confirmed that no other electron beam energy uses this scattering foil. Approximate radial and transverse coordinates were manually entered, and the “Drive To” function was used to position the carousel. To center the scattering foil and optimize beam symmetry, real-time beam profiles were acquired using the IC Profiler (Sun Nuclear Corp., Melbourne, FL, USA) under conventional dose rate. Incremental adjustments were made as needed, and the final coordinates were recorded to enable reproducible alignment in future experiments.

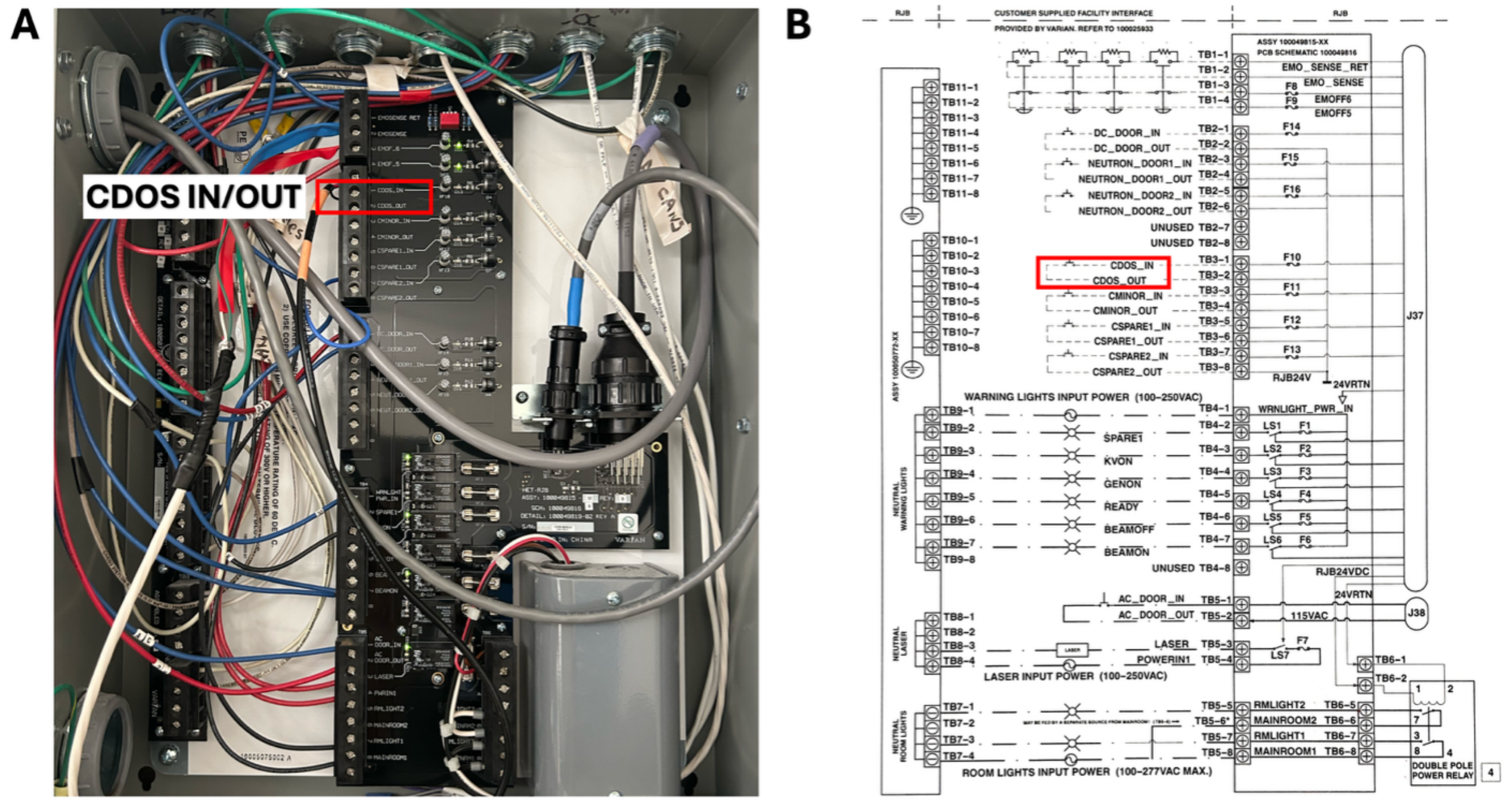

Figure S3: Relay Junction Box (RJB) located on a room wall. (A) Internal view of the RJB showing location of CDOS input and output sockets highlighted in a red box. (B) Corresponding RJB wiring diagram with CDOS input and output highlighted in a red box.

## S2.3 Determination of Coordinates for Removing Photon Target and Ion Chamber

The coordinates for photon target removal were determined by accessing the carousel interface while an electron energy was loaded. To identify the retracted position of the ion chamber, the system was placed in Idle mode, and the relevant coordinates were recorded from the interface.

## S2.4 Interlocks

Due to the manual rearrangement of the carousel, the setup bypasses the system's automated positional verification, and a series of beam generation and monitoring position (BGM.POS.) interlocks and slot mismatch interlocks are triggered. These interlocks must be overridden to allow beam delivery under manual positioning conditions.

- BGM.POS.energySwitch.NotAtPos
- BGM.POS.YAxis.NotAtPos
- BGM.POS.IonChamber.NotAtPos
- BGM.POS.Carousel.NotAtPos

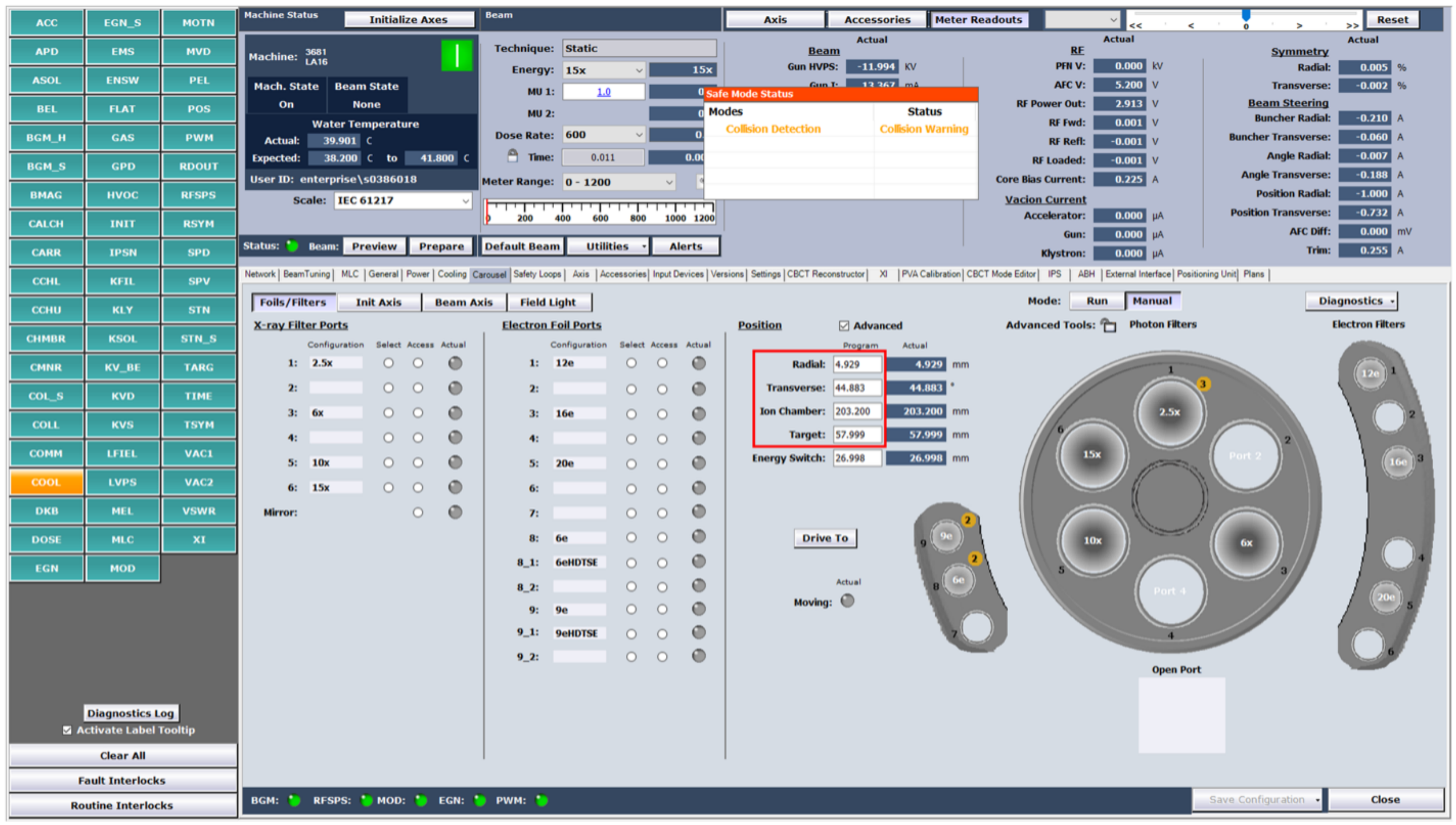

Figure S4: Carousel coordinates under the Carousel tab in Advanced Service Mode (default clinical operation values displayed). User can configure radial position, transverse position, ion chamber and target positions in the locations highlighted in a red box.

- BGM.POS.Target.NotAtPos
- BGM.POS.IonChamber.SecCheckDisabled
- BGM.POS.Carousel.SecCheckDisabled
- BGM.POS.Target.NotAtPos.SecCheckDisabled
- BGM.POS.YAxis.SecCheckDisabled
- BGM.SW.Axes.NotAtPosition

# S3 Beam Tuning

## S3.1 Determination of Maximum Dose Rate (performed once)

To establish the maximum dose rate of the 12 MeV FLASH, we first loaded clinical 15X photon mode. In Service Mode, at the Beam Tuning tab, the DOSE, automatic frequency control (AFC) and pulse forming network (PFN) servos were switched OFF, and the DOSE interlock was manually overridden. The high voltage power supply (HVPS) was set to 1.5 V below its default clinical operation value, in accordance with the recommended safety limits to avoid arcing. Next, the radio frequency drive (RFDR) voltage and the gun grid voltage were adjusted in parallel to account for beam loading and to maximize transmission through the bending magnet system. The goal is to increase the dose rate until saturation, which was observed when output readings reached approximately twice their conventional value (e.g., 15X at 600 MU/min saturated near 1200 MU/min). If this dose rate was not achieved, we reduced the electron gun's extraction pulse width (Gun (ext)) to decrease beam spread and enhance output. Once optimal values for RFDR, gun grid voltage, and HVPS were established, they served as the standard configuration for subsequent FLASH experiments. These parameters remained valid unless clinical service personnel performed a retuning of beam energies, in which case the maximum dose rate determination had to be repeated to reestablish optimal settings.

### S3.2 Experimental Setup (performed in each experiment)

After adjusting the carousel and overriding the relevant interlocks, beam tuning was performed by remaining in Manual Mode within the Beam Tuning tab. All Servos were switched OFF (DOSE, AFC, PFN, Angle R, Angle T, Pos R, and Pos T) to allow direct control of the beam parameters. The previously determined values for Gun (ext), HVPS, and RFDR were then entered to replicate the conditions established during the maximum dose rate tuning.

### S3.3 Reverting to clinical settings

To revert the system back to clinical mode, all Servos should be turned ON within the Beam Tuning tab, interlocks should be un-overridden, and the carousel should return to "Run" mode. The system then displayed a message indicating that the "BGM configuration has been modified" and giving the option to save changes. It is critical that these changes are not saved. To prevent accidental overwriting of the clinical configuration, we saved the clinical settings in advance so they could be restored if needed. As an additional safeguard, we have implemented in our experimental protocol daily quality assurance (QA) dose output checks before and after each experiment, and furthermore, there is a comprehensive daily QA prior to each day's clinical work. Therefore, if incorrect settings were mistakenly saved, the machine would generate interlock or configuration mismatch errors during post-experiment daily QA check.

## S4 Beam Control and Monitoring

The CDOS of the linac is activated by a high-speed reed relay controlled by a LabVIEW program through a compact configurable I/O (cRIO-9030; National Instruments, Austin, TX, USA). When the LabVIEW program is running, the relay closes and the beam is ready to be loaded. Once the beam is loaded, user needs to manually hold the beam and monitor the reflected radiofrequency (RF) power probed through the load power terminal. This allows the system to approach consistent resonance and stabilize RF output prior to beam delivery (*1, 2*). Once the reflected RF power reaches a stable waveform (Figure S5), user releases the beam.

Figure S6 shows the LabVIEW beam control graphical user interface (GUI) and a custom MATLAB pulse visualization GUI. The LabVIEW GUI allows users to define the number of pulses to be delivered, thereby setting the total dose. The dose per pulse is adjusted by adjusting the gun grid voltage in the service mode, with higher voltages corresponding to higher dose per pulse. The MATLAB GUI displays the acquired waveforms and measurements, including pulse charge (derived from ACCT readings) and pulse width. For each collimator configuration, film-based calibration is performed across a range of ACCT charges and monitor units (MUs). These calibration factors are stored and used during experiments to determine the target ACCT charge or MU required to achieve the prescribed dose. Additionally, these calibration factors can be entered into the MATLAB GUI, enabling real-time dose derivation from the ACCT charge during the experiments.

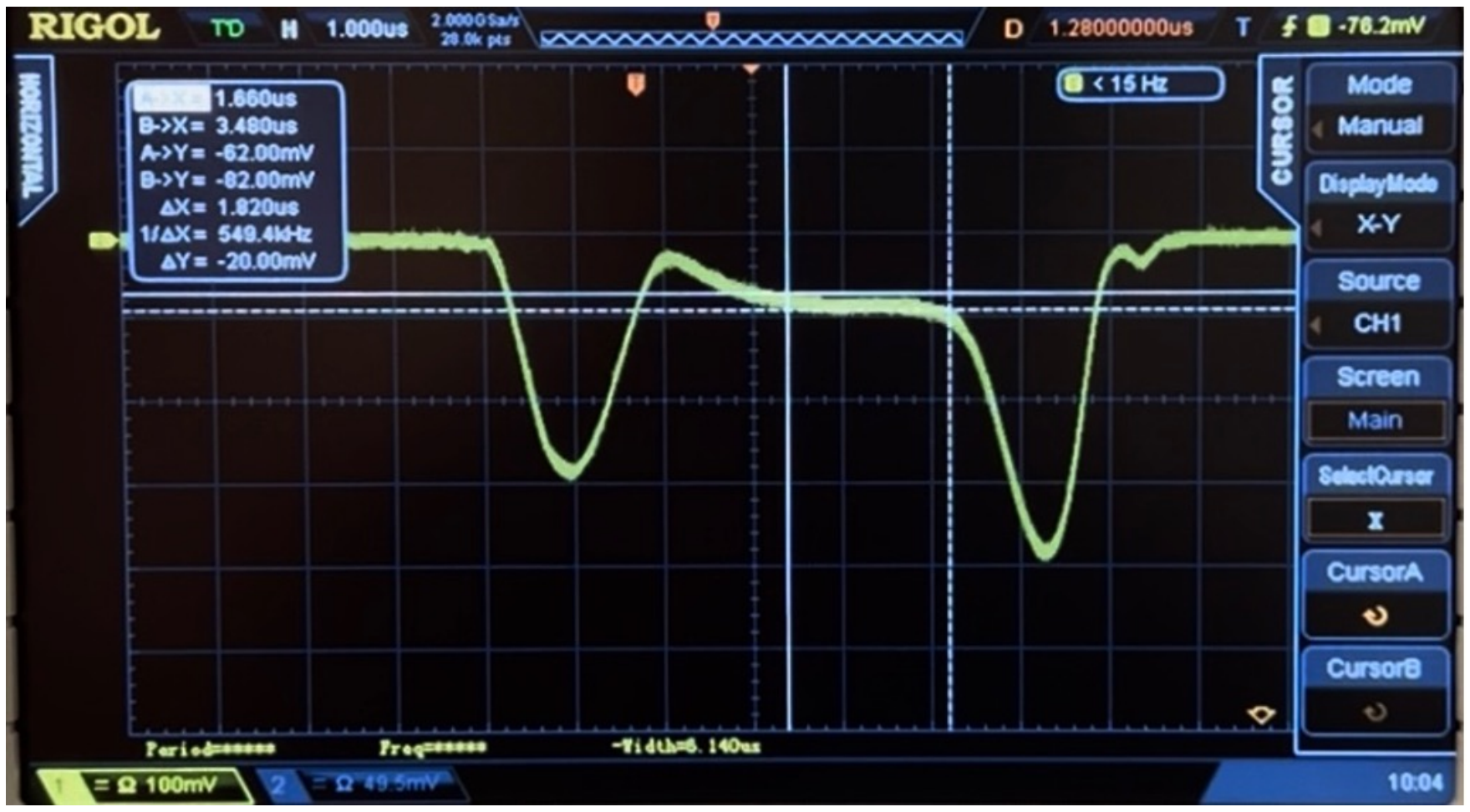

Figure S5: Example of a stabilized reflected radiofrequency (RF) power waveform, indicating readiness for beam release.

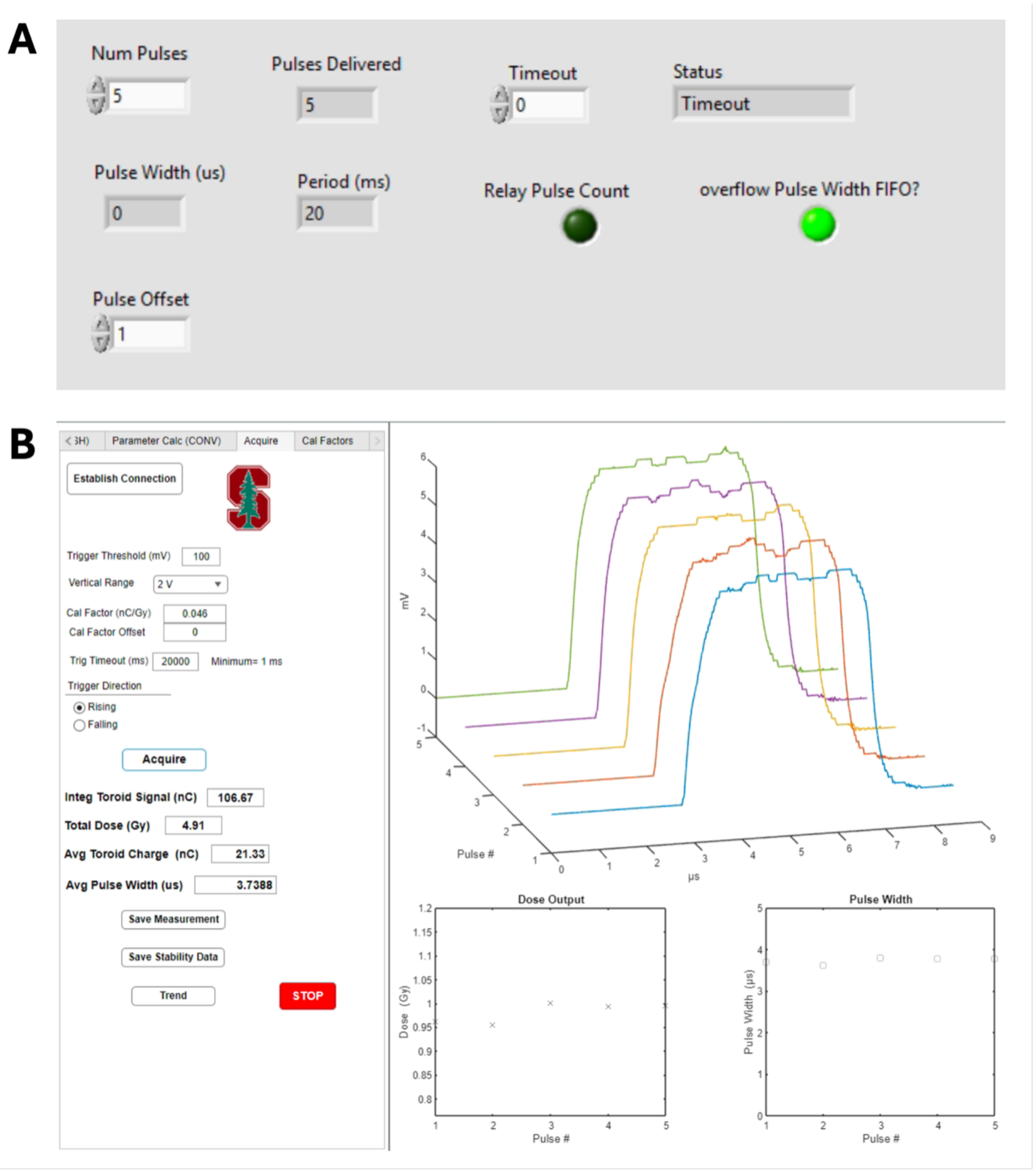

Figure S6: Example delivery of five FLASH electron pulses. (A) LabVIEW graphical user interface (GUI) used to control beam delivery and specify the number of pulses. (B) MATLAB GUI displaying AC current transformer (ACCT) charge readouts and pulse visualization.

# S5 Flatness and symmetry

Flatness was quantified as the relative difference between the maximum and minimum dose values within the central $d = \pm 2$ cm of the beam profile for the open field setup at isocenter, normalized to their sum:

$$\text{Flatness (\%)} = \frac{D_{\text{max}}^{(\pm d)} - D_{\text{min}}^{(\pm d)}}{D_{\text{max}}^{(\pm d)} + D_{\text{min}}^{(\pm d)}} \times 100. \tag{S1}$$

Symmetry was evaluated over the same central regions for the open field setup by calculating the maximum absolute difference between dose values at positions symmetric about the central axis, normalized to the central axis dose:

$$\text{Symmetry (\%)} = \max_{x \in |0,d|} \left( \frac{|D(x) - D(-x)|}{D_{\text{CAX}}} \right) \times 100, \tag{S2}$$

where $D(x)$ and $D(-x)$ are the doses at positions equidistant from the central axis, and $D_{\text{CAX}}$ is the dose at the central axis ($x = 0$).

The flatness and symmetry of the *in vivo* and *in vitro* setups were evaluated with the above equations at a 80% and 65% field width respectively.

# S6 Customized collimators and setups for *in vivo* experiments

To accommodate various research objectives, a suite of modular collimators and positioning jigs was developed (Figure S7). For *in vivo* studies, the system supports targeted irradiation of specific anatomical sites, including large-field abdominal and head/neck treatments (Figures S7A, B), as well as localized whole-brain and subcutaneous tumor geometries (Figures S7C, D). Specialized pulmonary setups enable both unilateral and bilateral lung irradiation, with the latter allowing for optional mediastinal shielding via a customized lead insert to protect the heart (Figures S7E, G). Additionally, the system includes a dedicated vertical frame for *in vitro* multiwell plate irradiation (Fig. 1H). All collimators utilize a 3D-printed shell filled with 1.2 cm of Cerrobend, providing robust shielding for both UHDR and CONV energy ranges.

# S7 *In vitro* setup

Figure S8 illustrated the wide-field irradiation setup and dosimetric performance at SSD beyond 80 cm for *in vitro* FLASH experiments.

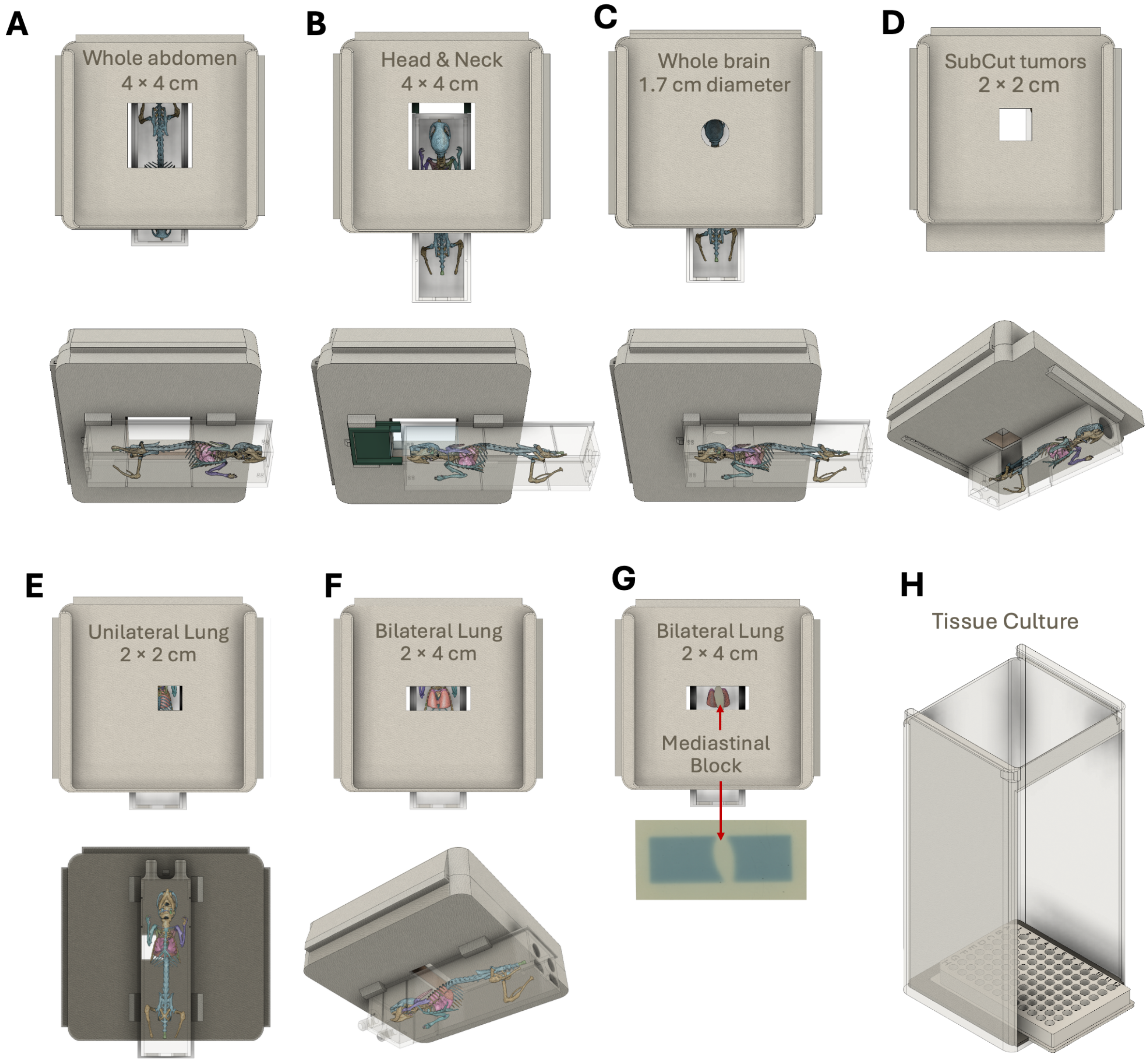

Figure S7: CAD renderings of customized collimators and irradiation setups used for preclinical *in vivo* and *in vitro* experiments. (A–G) Top and isometric views of animal positioning and collimator designs targeting specific anatomical regions: (A) whole abdomen ($4 \times 4$ cm$^2$), (B) head and neck ($4 \times 4$ cm$^2$), (C) whole brain (1.7 cm diameter), (D) subcutaneous tumors ($2 \times 2$ cm$^2$), (E) unilateral lung ($2 \times 2$ cm$^2$), (F) bilateral lung ($2 \times 4$ cm$^2$), and (G) bilateral lung with mediastinal shielding using a customized lead block. (H) Tissue culture setup for vertical irradiation of multiwell plates. All collimators consist of 3D-printed shells filled with 1.2 cm-thick Cerrobend for radiation shielding and collimation.

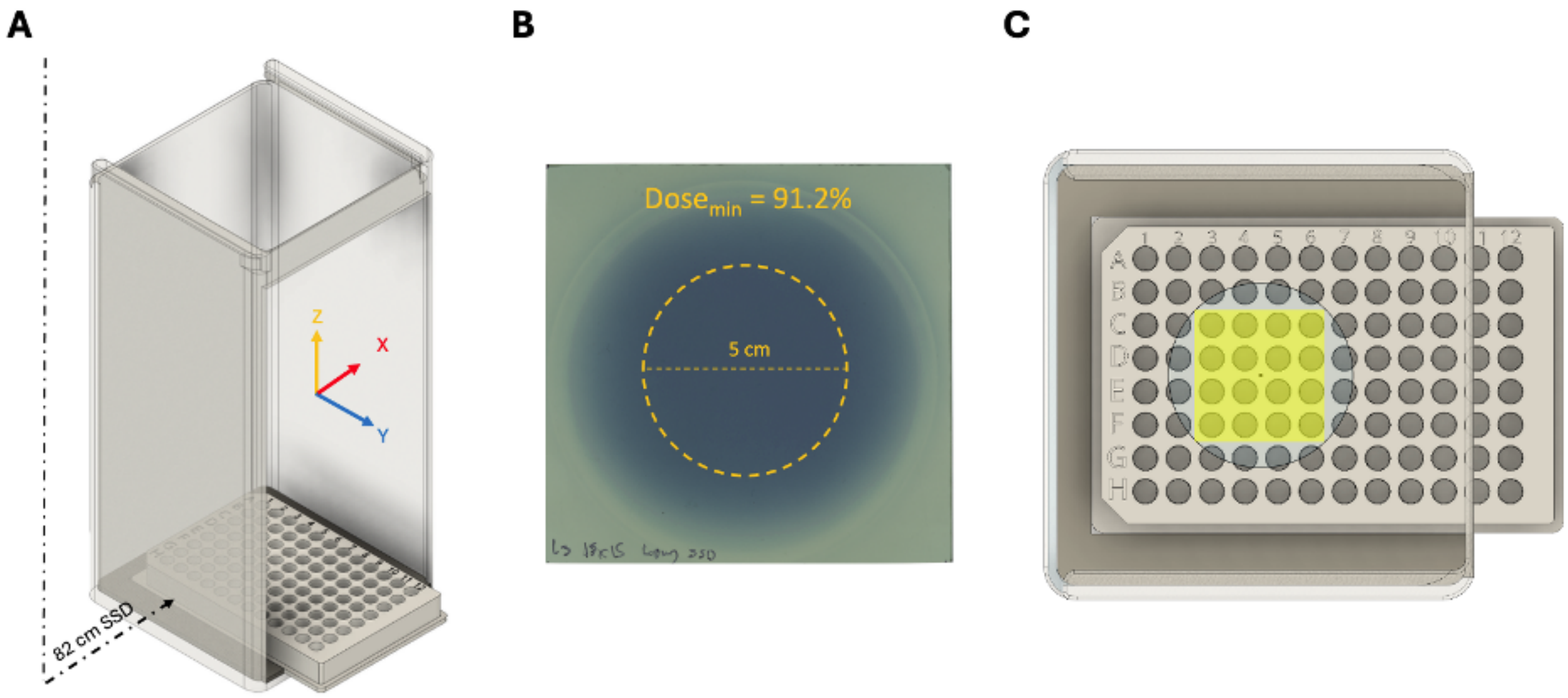

Figure S8: Wide-field irradiation setup and dosimetric performance at extended SSD for *in vitro* in FLASH applications. (A) CAD model of the irradiation chamber with a 96-well plate holder placed at 82 cm SSD. (B) Radiochromic film irradiated at 82 cm SSD under a 15 × 15 cm jaw opening using FLASH. A central 5 cm diameter region (dashed circle) demonstrates a minimum dose of 91.2% relative to the maximum, indicating acceptable flatness for uniform irradiation. (C) Top-down view of a 96-well plate showing the central region (yellow) where 16 wells fall within the area receiving ≥91.2% of the maximum dose. This setup supports FLASH-compatible irradiation for tissue culture and other wide-field applications.

# S8 Beam Parameters and Calibration Methods

We performed an *in vivo* five-day fractionated whole-brain irradiation experiment using the Truebeam FLASH platform. Table S1 shows the dose and beam parameters for the UHDR and CONV groups.

Table S1: Beam parameters for *in vivo* UHDR and CONV irradiations (target dose: 5 Gy, 6 Gy). Beam energy is the mean energy incident on the phantom surface derived from $R_{50}$.

| Mode | UHDR | CONV |
|---|---|---|
| Beam energy [MeV] | 12.8 | 11.9 |
| Target dose [Gy] | 5, 6 | 5, 6 |
| Delivered pulses | 5, 6 | 4017, 4795 |
| Pulse rate [Hz] | 180 | 90 |
| Dose per pulse [Gy] | 1 | $1.3 \times 10^{-3}$ |
| Dose rate [Gy/s] | 225, 216 | 0.11 |
| Pulse length [$\mu$s] | 3.2 | 3.2 |
| Intra-pulse dose rate [Gy/s] | $3.1 \times 10^5$ | 361 |

Prior to the experiment, calibration curves were established for UHDR and CONV respectively. For UHDR, the calibration curve was derived by delivering a series of doses at varying gun grid voltages to a 3D-printed anatomically realistic mouse phantom. The doses in the mouse brain at a depth of 0.9 cm were measured using films and the corresponding ACCT charges were recorded. A linear regression between measured film doses and ACCT charges was performed to establish the calibration curve. The target ACCT charge corresponding to the prescribed dose was then determined from the calibration curve, and the gun grid voltage was tuned to achieve the desired ACCT charge on the day of experiment. For CONV, the calibration curve was established by performing a linear regression between measured film doses and delivered monitor units (MU). The MU corresponding to the prescribed target dose was then selected for each irradiation. All calibration curves were specific to the irradiation geometry and collimator used.

During the experiment, ACCT charges were recorded for all FLASH irradiations and ion chamber charges were recorded for all CONV irradiations since the CONV dose rates fell outside the dynamic range of our ACCT. For each day of the experiment, quality assurance (QA) film measurements using the mouse phantom were acquired for each target dose, with five replicates per dose per fraction for UHDR and three replicates per dose per fraction for CONV. Two a posteriori calibration curves were derived by correlating QA film measurements with ACCT charges for UHDR and with ion chamber charges for CONV. These two curves were used to estimate the delivered dose for each mouse in the respective groups.

For the UHDR groups, the film dose correlated strongly with both ACCT charge and the downstream ionization chamber charge (Figure S9A), yielding linear regression fits with high coefficients of determination ($R^2 = 0.97$ and 0.96, respectively). These results confirm the linearity and consistency of dose estimation from both ACCT and downstream ionization chamber readings. For the CONV groups, the film dose was correlated with the downstream ionization chamber charge, which also demonstrated excellent linearity ($R^2 = 0.95$; Figure S9B), supporting its use as the primary dosimetric reference for *in vivo* dose estimation in the absence of ACCT-based readouts.

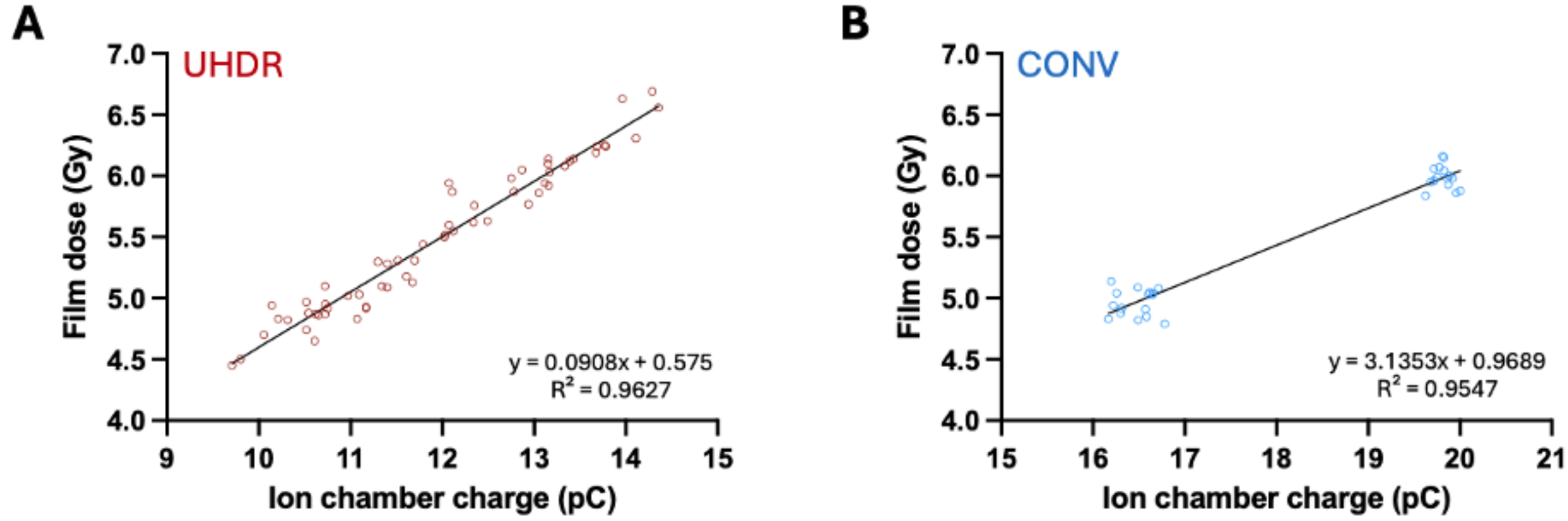

Figure S9: Calibration curves correlating external ion chamber charge with film-measured dose for FLASH and CONV. (A) Linear regression of film dose versus external ion chamber charge for FLASH, showing strong correlation ($R^2 = 0.9627$), supporting accurate dose estimation using real-time external ion chamber measurements. (B) Corresponding calibration for CONV, also demonstrating excellent agreement ($R^2 = 0.9547$). These results validate the use of external ion chamber charge as a reliable surrogate for delivered dose in both beam modalities when using the 3D anatomical mouse phantom.